\documentclass{article}
\usepackage{arxiv}
\usepackage{subfiles}
\usepackage[linesnumbered,ruled,vlined]{algorithm2e}

\usepackage{amssymb}
\usepackage{array}
\usepackage{url}
\usepackage{graphicx}
\date{}

\usepackage{soul}
\usepackage{xcolor}
\newcommand{\sg}[1] { \textcolor{brown}{{\bf SG: }{``\em #1''}}}

\newcommand{\puzzlers}[0]{puzzlers}
\newcommand{\puzzler}[0]{puzzler}
\newcommand{\Puzzlers}[0]{Puzzlers}
\newcommand{\ourname}[0]{\textsf{PUPoW}}
\newcommand{\vc}[0]{\textsf{VanityCoin}}
\newcommand{\ev}{Easy Verifiability}
\newcommand{\ad}{Adjustable Difficulty}
\newcommand{\pa}{Proportional Advantage}
\newcommand{\nr}{Non-reusable}

\newtheorem{proposition}{Proposition}

\SetCommentSty{mycommfont}
\SetKwComment{Comment}{$\triangleright$\ }{}

\SetKwInput{KwInput}{\textbf{Input}}
\SetKwInput{KwOutput}{\textbf{Output}}

\author{Yash Chaurasia \\
	International Institute of Information Technology\\
	\texttt{yash.chaurasia@research.iiit.ac.in} \\
	\And
	Visvesh Subramanian \\
	International Institute of Information Technology\\
	\texttt{visvesh.subramanian@research.iiit.ac.in} \\
    \And
    Sujit Gujar \\
	International Institute of Information Technology\\
	\texttt{sujit.gujar@iiit.ac.in} \\
}

\begin{document}
\title{\emph{PUPoW}: A Framework for Designing Blockchains with Practically-Useful-Proof-of-Work \\ \emph{\& VanityCoin}}
\maketitle

\begin{abstract}
Bitcoin is the first of its kind, a truly decentralized and anonymous cryptocurrency. To realize it, it has developed a blockchain technology using the concept of \emph{`Proof of Work'} (PoW). The miners, nodes responsible for writing transaction database, solve a cryptographic puzzle to claim the right to write to the database. Though bitcoin and many other relevant cryptocurrencies such as ether use revolutionary ideas, the main criticism involves the computing resource and energy consumption to solve the puzzles that have otherwise no use. There are attempts to use the PoW to do something useful, commonly referred to as \emph{Proof-of-Useful-Work} (PoUW). In this paper, we attempt to (i) make PoUW more usable -- describe how a central problem setter can crowdsource their work as PoUW and (ii) in the true spirit of blockchains, decentralize the role of problem setter, whom we call \emph{\puzzlers}. We propose a formal framework to do so, namely \emph{\ourname}. 
\ourname\ has an inbuilt provision of payments from \puzzler\ to the miner who solves its puzzle. Additionally, miners have the option to not rely on continuous feed of the puzzles and instead use original PoW puzzles. 

We also propose a way to use \ourname\ for solving TOR vanity URL generation and bitcoin vanity address generation problems. We call this \ourname\ blockchain solving vanity address generation problems as \vc. Both the problems require generating public keys from private keys such that resultant addresses are of interest. Such key pairs are found only by a brute force search. However, there are privacy concerns that miners would know the private keys of the puzzlers. We resolve this by splitting the private keys, and the miners would know only one part of it. In summary, we are proposing how PoW can be made practically useful, and we believe such an approach is needed for PoW blockchains to survive.\footnote{The preliminary version was presented at IEEE Blockchain 2021} 
\end{abstract}

\section{Introduction} \label{abstract}

Bitcoin: A Peer-to-Peer Electronic Cash System \cite{btc}, published in 2008, introduced us to \emph{blockchain} and Nakamoto consensus. A blockchain is a decentralized and distributed chain of blocks such that each block stores the hash of the previous block, thereby prohibiting any changes in the committed blocks once newer blocks are published. Bitcoin solved many problems in the online currency systems by using \emph{Proof-of-Work} (PoW) as the mechanism to achieve consensus. \emph{Miners}, who maintain the blockchain, try to publish blocks to earn \emph{mining rewards} by solving a computationally intensive cryptographic \emph{puzzle}. With PoW, bitcoin proved that eventual distributed consensus is possible under mild technical assumptions.  
\par
One of the key factors in PoW is that the probability of a miner mining a block is equal to the fraction of the entire network's computational power held by them. Consequently, miners keep investing in these resources as long as it is profitable to do so. Apart from establishing consensus in Bitcoin's network, the solutions to these puzzles have no use. These two factors combined mean that a lot of electricity (120 TWh per year \cite{digistat}) and hardware resources ($91\times10^{18}$ Hashes per second \cite{blockchaincom}) are spent just to mine Bitcoin, with no other useful work done. The total electricity used in mining bitcoin exceeds the electricity consumption of many countries \cite{digistat}, which has been a major criticism of Bitcoin. This criticism also had been one of the major reasons for the crypto-crash in May 2021~\cite{elontweet,yahoocryptocrash}, reducing the public trust in all cryptocurrencies in general, even those not relying on PoW for consensus. If the total hardware resources invested in Bitcoin mining also produce useful work, any supercomputer built in the foreseeable future won't parallel the network's capability. The implications will be profound depending on how useful the problem being solved is.

Currently, researchers are investigating if this computational intensive process of achieving consensus can be made useful. 
\emph{Primecoin} \cite{primecoin} requires miners to produce a Cunningham chain or bi-twin chain of some length decided by the computation power of the network as proof-of-work. Though these \emph{chains of primes} generated as PoW in Primecoin are available to everyone for use, its practical applications are limited.
\emph{PermaCoin} \cite{permacoin} replaces proof of work with proof of retrievability. The idea proposed here ensures that participating nodes store a very large data-set (\textit{like a library of books}) in a distributed manner.
\emph{Proofs of Work From Worst-Case Assumptions} \cite{pouwfromwca} gives proof-of-work based on problems like Orthogonal Vector, 3SUM, All-Pairs Shortest Path, and other problems which reduce to them.
\emph{A Proof of Useful Work for Artificial Intelligence on the Blockchain} \cite{pouwforai} describes a protocol that lets clients to submit Machine Learning tasks to the miners and pay all the actors involved in the process (miners, supervisors, and evaluators) if the model is satisfactory. Miners get to mine with some nonce after every iteration. The problem solved here is not easily verifiable and regular nodes only check if the hash is less than the network target $T$.
\emph{Proof of Learning (PoLe)} \cite{pole} is a consensus algorithm having data nodes and consensus nodes to do ML on blockchain. Data nodes push training requests and consensus nodes choose the request with highest reward/time value to train. The model with the highest test accuracy wins and the corresponding block is published.
\emph{Gridcoin} \cite{gridc} is a proof-of-stake blockchain that awards gridcoins to individuals contributing to BOINC white-listed crowdsourcing projects.
In \emph{Coin.AI} \cite{coinai}, miners must design a neural network instead of solving the hash puzzle. The previous hash, the nonce, and the list of transactions determine a neural network architecture. A miner trains this model on a publicly agreed-upon data-set. Any model which crosses a threshold is valid.
These works are often referred to as \emph{Proof-of-Useful-Work} (PoUW), the work that is used for two purposes: (i) to solve some useful problems, and (ii) to prove that some problem is solved to use as PoW to maintain a distributed ledger.
\par

The PoUW-based blockchains are limited at this point as  (i)  they propose a very specific problems which leads to solving limited types of useful problems. (ii) Some proposals involve centralized administrator and data sharing (e.g., Coin.AI or Permacoin) with the miners. (iii) A currency built on top of such blockchain should continue to be useful forever. At this point, it is not clear that in the absence of any problem to be solved how these blockchains will continue. 
There is a need for a formal approach to build a PoUW protocol that leverages  PoW concepts and the work done for achieving consensus on the blockchain has some usefulness in the real world. 

\textbf{Contributions:} In this work, we list important characteristics of the problem that one can use to build a PoUW blockchain, namely, \emph{\ev, \ad, \pa,} and \emph{\nr}. These essentially capture the fact that the problem should be hard to solve but a solution should be easy to verify, we should be able to adjust the difficulty easily, the chance of solving should be proportional to the computing power, and the solution to one problem in a previous block cannot be reused in another block.  With these, we propose a framework in Section \ref{centralizing} to build a PoUW based blockchain with a center, whom we call \emph{problem setter},
that sets up the problem to be solved. However, in the true spirit of decentralization, it is better to have a decentralized system where anybody can participate as a problem setter. In decentralized PoUW, we refer to them as \emph{\puzzler}, provided the problem they set meets the desired conditions. Towards this, we propose a framework, \ourname, a decentralized PoUW blockchain framework. Key interesting features of our framework are (i) anybody can be a \puzzler, (ii) in the absence of problems to solve, it can revert to a normal PoW blockchain, (iii)
provision of \puzzlers\ rewarding miners (thereby reducing the importance of transaction fees). We illustrate how to use \ourname\ with an interesting problem, specifically, \emph{generating vanity .onion URL}.
One can use a similar technique for generating vanity addresses in bitcoin and other applications using DSA-like scheme to generate addresses. We refer to a \ourname\ blockchain generating vanity addresses for proof-of-work as \vc.

\section{Notations \& Definitions} \label{notations}
\subsection{Bitcoin's Proof-of-Work Puzzle}
An online decentralized currency needs a mechanism to achieve consensus considering that the system will have some malicious participants. Satoshi Nakamoto brought a revolution in online currency systems by introducing \emph{Proof-of-Work} protocol which achieves consensus in the network and is also safe against attackers having less than 50\% of the network's computing power. Miners get to accept and publish valid transactions on their blocks but to publish their block they need to solve a \emph{partial hash pre-image puzzle}, i.e., the hash of their block header should be less than some target. \\
$$ \mathcal{H}(nonce || prevhash || merkleroot || timestamp || version || target)<t $$
The target $t$ is adjusted after every 2016 blocks so that the average block publishing rate remains $\sim$1 block per 10 minutes. To try out different hashes, the miners change the nonce field. The block header also has a field for the hash of the previous block's header, also known as \emph{hash pointer}. 
We discuss components of Bitcoin-like blockchains next.
\subsection{Blockchain Components}
\begin{itemize}
    \item \emph{Blocks}: These contain the transactions in a Merkle tree and the block header.
    \item \emph{Block header}: This contains the hash of the previous block, Merkle root of the transactions, timestamp, nonce, version and target.
    \item \emph{Transactions}: Each transaction contains inputs and outputs. Inputs typically have signatures and public keys. Outputs have scripts depending on the usage.
    \item Merkle Tree: All the transactions are contained outside the block header of the block and arranged in a Merkle tree. The Merkle root is committed in the block header.
    \item Transaction mempool: Valid transactions generated by the users are relayed to other users/miners. Transaction mempool is where valid transactions wait to be confirmed.
\end{itemize}

\subsection{Hash function $\mathcal{H}$}
A hash function $\mathcal{H}$ takes an input $i$ and outputs hash $h$ with $n$ bits. For our purpose, the input will be the block header of the block. We write this as $\mathcal{H}(block\_header)$, where $$block\_header =  (nonce || prevhash || merkleroot || timestamp || version || target)$$ in the case of Bitcoin. It is possible to build a hash function outputting hash of any required length $n$.
\par

\subsection{Actors in our framework}
\paragraph{Miners:} Actors who solve the problem to be able to publish their block are called miners. They receive fixed mining rewards for publishing a block and a variable problem and transaction fee based on the problem/transactions chosen.
\paragraph{Users:} Full nodes and Light nodes which only focus on transactions happening on the blockchain are called users.
\paragraph{Problem setter:} In the centralized scenario, the authority responsible for setting the problem for the miners to solve is called the problem setter.
\paragraph{\Puzzlers:} In the decentralized scenario, the actors who propose their problems to be solved are called \puzzlers.


\subsection{Practically Useful Work}
%
A lot of effort has been put to identify useful algorithmically generated problems, but limited success has been achieved. We thus define a practically useful problem to be one that is proposed by an individual/organisation, a problem for which the individual/organisation might be willing to pay to get solved.
For example, the primecoin protocol won't be considered as a practically useful PoW under this definition, but if a \puzzler\ requests to get a Cunningham chain, this will be considered a practically useful problem.
\emph{Practically-Useful-Proof-of-Work} (\ourname) framework shows how such practically useful problems can be used in place of Bitcoin's puzzle as Proof-of-Work.

\subsection{Important Problem Requirements}
For any problem to be useful in building a PoW blockchain, the following properties are essential. \cite{IEEEexample:bitcoin-book}:
\paragraph{Easy-verifiability:} A solution to a problem must be easily verifiable by anyone, and verification time should not vary much with the difficulty of the problem. In Bitcoin, the \emph{partial hash preimage} puzzle is easily verifiable, although a preimage can only be found by brute force.
\paragraph{Proportional advantage:} The probability of a miner to mine a block must be proportional to their computational power relative to the network's power. For this, the problem being solved must be progress free and memoryless, i.e. no advantage is gained by previous attempts at finding a solution, and each attempt is equally probable to succeed. Thus, the no. of attempts (or indirectly, the computational power) of a node is the only factor relevant for success. In Bitcoin, as the solution has to be brute-forced and because the hash is considered to be random, all the properties are satisfied.
\paragraph{Adjustable difficulty:} Similar to bitcoin, to ensure some security aspects, the rate at which blocks are being published needs to be controlled. For this, the difficulty of the problem being solved is adjusted based on the computing power of the entire network. In Bitcoin, the objective of the puzzle is to get a preimage such that the hash of the block header is less than some target. This target can be increased or decreased to adjust the difficulty of the problem.

Apart from the above properties from \cite{IEEEexample:bitcoin-book}, we also list the following important property a PoW protocol should have to be useful. 
\paragraph{Non-reusable:} Any solution in one block should not be reusable in another block, or else this can then be used to do cheap work, and the block publishing rate will no longer be controllable.

\subsection{Other Problem Requirements}

Some of the other requirements discussed in \cite{IEEEexample:bitcoin-book} are:
\paragraph{Equiprobable solution space:} All inputs that can be given are equally likely to produce a valid solution. This is essential to maintain progress-freeness because if the solution space is skewed, fast miners can try out promising inputs first.
\paragraph{Inexhaustible solution space} The blockchain network should not run out of problems to solve.

\paragraph{Algorithmically generated:} Blockchains often try to generate the problem algorithmically without having anyone to propose the problem. We identify this as one of the barriers to getting useful work done, so our proposal doesn't follow this.

\paragraph{ASIC resistance:} The idea that custom-built hardware specialized for mining shouldn't be much more efficient than a general-purpose computer so that it is profitable to mine on a general-purpose computer too.
 \\
\par
As mentioned earlier, we consider two scenarios - one where a central authority controls which problem is to be solved by miners and the other in which \puzzlers \ propose problems, like transactions, with the problem fee they are willing to pay for it. We discuss the centralized scenario in Section III and the decentralized scenario in Section IV.

\section{A Centralized Problem Setter} \label{centralizing}
Let there be a centralized problem setter $\mathcal{C}$. $\mathcal{C}$ broadcasts the problem to be solved, and miners need to solve the problem to mine the block. As the hash of the previous block is a part of the valid input to the problem, the problem can be made available in advance (as is in bitcoin - the target is known). We start describing the framework with a simple example of a real-valued univariate function.

\subsection{Real Valued Univariate Function}
\noindent \textbf{\textit{Problem:}} Let $f()$ be a real valued function in one variable such that for certain values of $x$, $f(x) < y_0$ for some real value $y_0$. $\mathcal{C}$ is interested to find some $x_i$ for which the condition is satisfied, i.e., $f(x_i)$ is less than the target value $y_0$. $f()$ is not assumed to have any exploitable property.
\\
\textbf{\textit{Protocol:}} Treat the 32 bit output $h$ from $\mathcal{H}$ as a integer/float value (similar to that of C programming language). Calculate $f(h)$ and check whether $f(h)$ is less than the threshold value $y_0$. If yes, the problem is considered solved, and the miner can proceed to publish their block. If no, the miner proceeds to find a new $h$ by picking a different nonce and repeating the above steps. Other miners/users can simply verify if the publisher solved the puzzle by computing $\mathcal{H}$ of the block and checking if the hash $h$ does in fact give $f(h) < y_0$.

\begin{algorithm}
\SetAlgoLined
\SetKwRepeat{Do}{do}{while}
 \Do{$f(h) >= y$}{$h=\mathcal{H}$(block header)}
 \KwResult{$h$ is a solution to the problem }
 \caption{Solving a univariate problem}
\end{algorithm}

\par
As long as $f(x)$ can be computed efficiently for all values of $x$ in approximately the same time, the solution will be easily verifiable as the hashing step can be done quickly. It will also give a proportional advantage to miners as they have to try out different nonces, and there is no progress if a hash fails to satisfy the condition, i.e., each nonce is equally likely to succeed in expectation.
\par
The difficulty of the problem depends on the function and the target. We assume $\mathcal{C}$ to have an estimate of how much time the network will require to publish a block for a target $y_0$. In case $\mathcal{C}$ requires a much lower $y_0$ than is suitable for the blockchain, $\mathcal{C}$ can set two targets - $y_1$ for the regular functioning of the blockchain and $y_0$ for which the miner can be awarded additional reward. The same problem can be set again until a block achieves $f(h) < y_0$ as the output. \par
This can be extended to solve a variety of problems. We now show how to extend this idea to multivariate functions.

\subsection{Multivariate Function}
\noindent \textbf{\textit{Problem:}} Consider $y = f^n(x_1, \ldots, x_n)\in \mathbb{R}$. Let the acceptable subset of the range given by $\mathcal{C}$ be $\mathcal{Q}$. Let $\mathcal{H}$ give $k * n$ bit output $h$.
\\ \textbf{\textit{Protocol:}} Divide the hash of the block $h$ into $n$ chunks of size $k$ bits such that their concatenation is $h$. Alternatively, this can be thought of as representing $h$ in $2^k base$, i.e. $h = x_1*(2^k)^0 + x_2*(2^k)^1 + \ldots x_n*(2^k)^{n-1}$. Use these $x_i$'s as input for the function $f^n$ and check if the output belongs to $\mathcal{Q}$. If yes, the problem is considered solved, and the miner can proceed to publish their block. If no, the miner proceeds to find a new $h$ by picking a different nonce and repeating the above steps. Other miners/users can simply verify if the publisher solved the problem by computing $\mathcal{H}$ of the block and checking if the hash $h$ and the resulting $x_i$'s do in fact give $f^n(x_1, \ldots, x_n) \in \mathcal{Q}$.
\begin{algorithm}
\SetKwRepeat{Do}{do}{while}
 \Do{$f^n(x_1,x_2,...x_n)\not\in \mathcal{Q}$}{ $(x_n||x_{n-1}...||x_1) = h =\mathcal{H}$(block header)}
  \KwResult{($x_1,x_2...x_n$) is a solution to the problem }
 \caption{Solving a multi-variate problem}
\end{algorithm}
\par
Same conditions hold true here as for the above scenario. With this step, it might be possible to do (randomised) linear regression by searching for vector $\alpha \in \mathbb{R}^p$ such that $x_i^T\alpha \approx y_i\; \forall i \in \{1, \ldots, n\}$ or $$l^* = \mathcal{L}(\alpha) = \frac{1}{n}\sum_{i=1}^{n} (x_i^T\cdot\alpha - y_i)^2$$ such that $l^* < l_0$ and $(x_i, y_i) \in \mathcal{D}$, the dataset. $$\mathcal{D} = \{(x_1,y_1), \ldots,  (x_n,y_n) : x_i \in \mathbb{R}^p \}$$
\par
This can be extended to set neural network weights too. This, of course, is sub-optimal for these applications, but better applications can be found. Functions that can't be exploited by techniques of gradient descent, etc., are good candidates.
\par
If $\mathcal{C}$ is trusted, $\mathcal{C}$ can also publish a black box code that takes input the hash of the block and outputs whether the condition is satisfied or not.
\subsection{Other Problems}
A central authority crowdsourcing science projects can adopt blockchain to distribute rewards and tie the problem to proof of work simultaneously. As the mining rewards are defined by the protocol, $\mathcal{C}$ cannot mint currency, limiting their control only to problem-setting. We consider an example problem to demonstrate how this can be done.
\\ \\
\textbf{\textit{Problem:}} High-resolution images from a telescope have been released. The task is to find supernovae. The data has been divided into $2^k$ small sectors. Miners are informed of how to spot a supernova (either manually or computationally).
\\
\textbf{\textit{Protocol:}} Let hash $h$ from $\mathcal{H}$ be of $k$ bits. Treat $h$ as an integer value - the sector to be searched. Look for a supernova in that sector; if one is found, then the miner gets to publish the block; else, try another sector with a different nonce. Others can verify by just looking at the sector given by the hash of the block. The miner can be rewarded more if a rarer object like, for instance, a magnetar is discovered.
\par
Similarly, projects looking for gravitational waves (e.g. Gravity Spy \cite{gravityspy}), planets in other star systems (e.g. Planet Hunters \cite{planethunters} \cite{planethunterstess}), etc. can tweak and adopt this model. Changing the difficulty might be tough here, but the relationship between the overhead for every sector and sector size can be looked into to enable changing of difficulty by changing the size of each sector. Often such projects use simulated data to train and test the citizen scientists. This can also be used to adjust difficulty.

\section{Decentralizing Problem Proposal} \label{decentralizing}

\begin{figure}
    \centering
    \includegraphics[width=0.85\columnwidth]{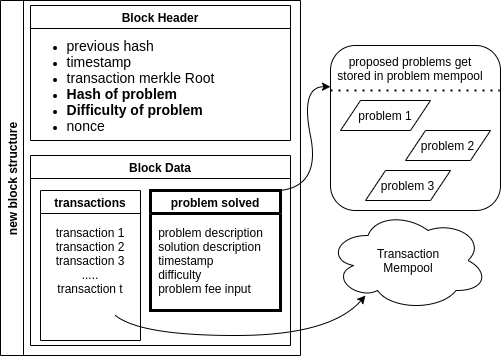}
    \caption{Block structure}
    \label{fig:block_structure}
\end{figure}

So far, we have assumed that the centralized problem setter $\mathcal{C}$ is trustworthy and can't be attacked by malicious parties. $\mathcal{C}$ cannot mint coins according to their wish but can either deny service or not follow the difficulty considerations, allowing them to control the frequency at which the blocks get published. Denial of service can be tackled by having an algorithmically generated problem to fall back to. Selective denial will be tougher to deal with and can be used to make various attacks.
\par
These factors necessitate coming up with a decentralized version of doing useful work. We started off by defining a practically useful problem as a problem requested by someone for this reason - anyone can propose their problem to be solved by offering a problem fee. We call these people/actors as \emph{\puzzlers}.
\par
Like users propose transactions, \puzzlers \ propose their problem. There is a problem mempool similar to transaction mempool. \Puzzlers \ need to put timestamp, the difficulty for which the problem was proposed, signature, the fee offered, etc. We first describe the additional components and the new fields in the block and the block header required to enable \ourname. 

\subsection{Additional Components of the Blockchain}
\begin{itemize}
    \item \emph{Block}: This additionally contains the problem that the miner selected to solve.
    \item \emph{Block header}: This additionally contains the Merkle root of new problems, hash of the problem, and instead of the target, it contains the difficulty of the problem solved.
    \item \emph{Problem Proposal}: Contains problem description, timestamp, the difficulty for which the problem was proposed, problem fee input and solution description (e.g. regex $\mathcal{R}$ in our proposed useful problem).
    \item \emph{Problem mempool}: Valid problem proposals are relayed to other miners. Once the difficulty is changed, older problems no longer remain valid. Problem mempool is where valid problems wait to be solved by miners.
\end{itemize}    

\begin{figure}
    \centering
    \includegraphics[width=0.85\columnwidth]{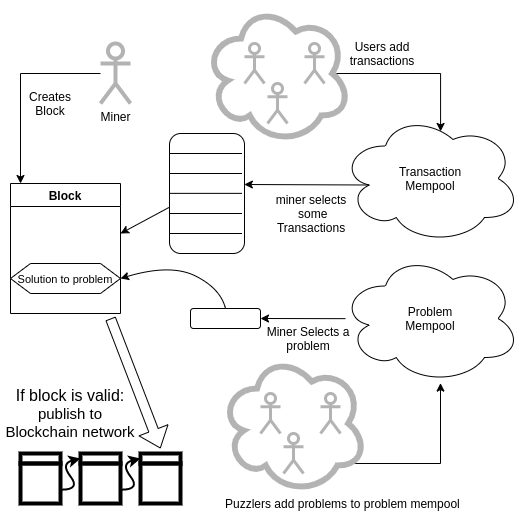}
    \caption{Decentralized environment}
    \label{fig:decentralized_environment}
\end{figure}

\subsection{Differences from the Centralized scenario}
In the centralized scenario, we assumed $\mathcal{C}$ to have an estimate of how much time the network will take to solve the problem. $\mathcal{C}$ was incentivised to maintain the blockchain and hence maintain the time period between blocks. In the decentralized scenario, the difficulty of the problem needs to be quantifiable or else malicious miners can propose easy problems themselves and get mining rewards faster. This means our problem space is restricted to problems whose difficulty can be easily characterized. More research on difficulty metric can allow this space to grow. Primecoin \cite{primecoin} showed how this is done for their problem.
\par
Let $\mathcal{S}_s$ be the set of all problem types $\mathcal{P}_i$ satisfying the above conditions. Thus, $\mathcal{S}_s = \{ \mathcal{P}_1, \mathcal{P}_2, \ldots \}$. Note that Bitcoin's cryptographic puzzle, $\mathcal{P}_b \in \mathcal{S}_s$. Let $\mathcal{S} \subset \mathcal{S}_s$ be the set of types of problems that can be proposed for PUPoW on a particular implementation of the blockchain. Details regarding rewards how to characterize the difficulty of each problem type is pre-defined. \Puzzlers \ can select a problem type $\mathcal{P}_i$ and ask for a specific instance satisfying the current difficulty of the blockchain. This is briefly shown in Algorithm \ref{alg:propose}.
For example, if the problem defined by Primecoin $\mathcal{P}_p \in \mathcal{S}$, a \puzzler \ can ask for a bi-twin chain of the length determined by the difficulty. Algorithm \ref{alg:solve} shows how a miner picks and solves a problem.

\begin{algorithm}
\SetAlgoLined
pick $\mathcal{P}_i \in S$ \;
$d_i =$ current difficulty for $\mathcal{P}_i$\;
design problem $p$ of type $\mathcal{P}_i$ for difficulty $d_i$\; \Comment{Add problem description, signature, solution description, problem fee input \& timestamp}
Broadcast $p$ to problem mempool\;
 \caption{\texttt{PROPOSE}: Propose a Problem} \label{alg:propose}
\end{algorithm}

\begin{algorithm}
\SetAlgoLined
select $\mathcal{P}_i$s where $\mathcal{P}_i\in S$\; \Comment{These $\mathcal{P}_i$s are problem types for which the miner has hardware to solve, $\bigcup_i\mathcal{P}_i \subseteq \mathcal{S}$}
Refresh\_Problem\_Mempool()\;
\Comment{gather problems for the current difficulty}
from problem mempool pick $p \in \bigcup_i\mathcal{P}_i$ \;
\SetKwRepeat{Do}{do}{while}
 \Do{New\_Block\_Published() == false}{$h=\mathcal{H}$(block header)\; \If{Solve($p,h$)}{Publish\_Block()\; break()\;}}
 Update\_Blockchain()\;
 \caption{\texttt{SOLVE}: Pick and Solve a Problem} \label{alg:solve}
\end{algorithm}

\par
We now describe one type of problem $\mathcal{P}_v$ satisfying the above criteria, i.e. $\mathcal{P}_v \in \mathcal{S}_s$.

\section{Generating Vanity .onion URL} \label{tor}

\subsection{Onion Addresses}
An onion address is 56 characters long for version 3 onion services (16 characters long for version 2), with characters being in base32 (i.e. 32 possibilities for each character, 2-7 or a-z). When a server is being set up, a public key - private key pair is generated by the Ed25519 signature scheme, and the public key, appended by the checksum and the version, is used as the address.
\par
For ease of exposition, we demonstrate our idea assuming DSA signing scheme instead\footnote{We discuss the exact details about how to use this with Ed25519 signature scheme in Appendix}.
\begin{itemize}
    \item $q$ is a prime.
    \item $p$ is a prime such that $q$ divides $p-1$.
    \item $h$ is an integer in $\{2, \ldots, p-2\}$.
    \item $g = h^{(p-1)/q}$ mod $p$.
    \item The parameters of DSA are $(p,q,g)$.
    \item private key: $x$ such that $x \in \{1,\ldots, q-1\}$
    \item public key: $y$ such that $y = g^x$ mod $p$.
\end{itemize}

\subsection{Vanity Addresses}
The addresses resulting from the above process are generally not human-readable, and often, organizations want their addresses to be more memorable and identifiable. Thus, they try out many key pairs to pick the best address suiting their requirements. This address is called a vanity address and can only be generated by brute force.
\par
This problem is thus suitable as \ourname\ if a way is found to adapt it for blockchains. We can ask miners to generate key pairs such that the hash of the block gives the private key and the public key needs to satisfy the \puzzler's requirement, but this would give away the private key to the entire network once the block is published and thus the public key would no longer be useful.
\subsection{Adapting the problem for blockchains}
\subsubsection{Problem Proposal}
Each character has 32 possibilities, and the difficulty is exponential in the number of characters required to be fixed. Miners, while mining, can come across some addresses that are preferred to the puzzler but couldn't be specified by fixing the characters. Thus, we allow \puzzlers \ to specify their acceptable addresses in the regex format. The difficulty of the regex expression can be calculated. Regex allows fine control over the difficulty. Instead of having a single difficulty in the blockchain specification, the blockchain gives a range of valid difficulties that the problem proposal can have.
\subsubsection{Keeping the private key secret}
As mentioned earlier, generating $x$ on blockchain to get suitable $y$ would not work. We need to generate $y$ without disclosing $x$ to miners. We do this as follows:
\begin{itemize}
    \item The \puzzler\ picks a random number $x_0$ and gives the value of $y_0 = g^{x_0}$ mod $p$ in the problem proposal.
    \item Hash $h$ of the block header, which also belongs to the private keyspace, is computed by the miner.
    \item $y = y_0 \times g^h$ is computed for $h$, and if $y$ matches the regex $\mathcal{R}$, the block can be published, or else, the miner tries a different nonce.
    \item Thus, $y = g^{x_0} \times g^h = g^{x_0+h}$. Therefore, the private key is $x = x_0 + h$.
    \item The blockchain network only knows $h$ while $x_0$ is known only to the \puzzler. Thus, $x$ is known only to the \puzzler.
\end{itemize}

\begin{proposition}
The miners or any users knowing the private key of the \puzzler\ corresponding to its vanity .onion url is as hard as solving discrete log problem.
\end{proposition}
\textbf{Proof:} If a \puzzler's .onion URL is generated, the miners and the users only know $y, h, y_0$. To know the private key, the attacker needs to know $x_0+h$ such that $y=g^{x_0+h}=y_0*g^h$. Thus, to know the private key, the attacker needs to solve either $y = g^{x}$ or $y_0=g^{x_0}$, i.e., solve a discrete log problem. 

\subsubsection{Verification}
Miners and users verify the block as follows:
\begin{itemize}
    \item Verify whether the difficulty of the regex $\mathcal{R}$ of the problem selected is within the current difficulty bounds of the blockchain.
    \item Compute hash $h$ of the block header.
    \item Calculate $y = y_0 \times g^h$
    \item Check if $y$ matches $\mathcal{R}$.
    \item Verify transactions, Merkle tree, etc.
    \item If any of the above fails, the block is considered invalid; else, the block is valid.
\end{itemize}
Thus, the private key is $x = x_0+h$ and the public key is $y = g^x$.
\subsection{VanityCoin}
We have primarily talked about .onion addresses here because a lot of energy is spent by organizations to generate them. For example, facebook used 500,000 cores and used 100,000 USD of electricity to generate their .onion address \cite{fb-onion}.
However, other applications involving similar public key/address protocols using DSA-like schemes can be used in \ourname. Bitcoin vanity addresses can also be generated similarly. Users often generate new addresses to improve their anonymity. We dub a \ourname\ blockchain with these problem types (optionally with $\mathcal{P}_b$, bitcoin's puzzle) in its problem type set $\mathcal{S}$ as \vc.

\begin{figure}[h]
    \centering
   \includegraphics[width=10cm]{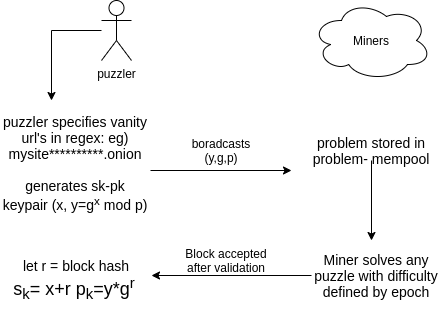}
    \caption{\vc\ in brief}
    \label{fig:tor_add}
\end{figure}



\section{Analysis} \label{analysis}
\subsection{Problem Requirements}
We have used the hash $h$ as the input to all the problems. The task of miners is to find $h$ for which some condition is satisfied, and they typically iterate over many nonces to be able to mine a block. This enables us to satisfy two major requirements, namely easy verifiability and proportional advantage. Verification of a block only involves hashing the block header, running the instance of the problem with the hash as the input (which, under our assumption, can be done efficiently) and transaction verifications. Given that miners need to use only the hash of the block as input and are supposed to try out different nonces, there will be no progress once a nonce fails, and as the hash is assumed to be random, every miner will have a probability of succeeding equal to their fraction of the computing power of the network.
\par
Adjusting difficulty might be easy in most problems, but interpreting a difficulty in terms of time required to solve it might be tough in many problems. Increasing the threshold $y_0$ decreases the difficulty, but it may happen that the difficulty changes non-linearly. We argued that in the centralized scenario, the central problem setter $\mathcal{C}$ needs to have an estimate of the difficulty of the problem they are proposing. Given that the network solves the problem for them, they are incentivised to maintain the block publishing rate. Like Coin.AI \cite{coinai}, a dynamic threshold can be explored. As mentioned earlier, some citizen science programs also insert simulated data to train the volunteers so the frequency of simulated data can be adjusted according to the difficulty. In the decentralized scenario, puzzlers propose problems, and if difficulty conditions are not strict, a miner might propose an easy problem and solve it themselves to get the block reward. Thus, characterizing difficulty is more important in the decentralized scenario. Our problem, generating vanity .onion address, and Primecoin's chain finding problem are two examples where this is shown to be possible. Further research will result in more such problems.
\par
It is often debated that ASIC resistance is a useful property that a blockchain puzzle can have. Many blockchains have tried to achieve ASIC resistance by having the miners solve various types of hashes to make it difficult to design ASIC. By having a set of allowed problem types for PUPoW, our design can enable this. Miners in the decentralized scenario might still be able to choose only a specific problem type if they want to, but the problem fee for a problem type with more miners will be expected to be lower. Thus, miners who can solve as many problem types as possible will have an advantage as they can adjust according to the demand. A blockchain network can vary the difficulty of individual problems separately too to bridge the gap between the efficiency of ASIC and personal computer (assuming certain problem types are ASIC friendly and some are PC friendly).
\par
The condition that the hash of the block header has to be the input removes the need to have an \emph{equiprobable solution space} (one attack is possible here, we discuss it in the next section). Skewed spaces can't be exploited as the miner has no choice over the part of space they search in. \\
We can safely assume that $\mathcal{C}$ or puzzlers can generate as many problems as the system requires. Also, Bitcoin's puzzle $\mathcal{P}_b \in \mathcal{S}_s$, can also be included in $\mathcal{S}$ but with no additional problem fee (as this won't be proposed by any \puzzler).
Thus, this satisfies \emph{inexhaustible solution space}. We deliberately move away from \emph{algorithmic problem generation}; hence our idea doesn't satisfy it.
\subsection{Attack surface}
Our framework is similar to Bitcoin, and there are only a few attack possibilities introduced on top of Bitcoin's vulnerabilities.
\subsubsection{Denial of Service}
In the centralized scenario, $\mathcal{C}$ broadcasts the problem to every miner. As discussed earlier, $\mathcal{C}$ can deny service at any time or be attacked by hackers who try to disrupt the blockchain. $\mathcal{C}$ can also selectively deny service to certain users, which can be used as a sanction against a group of people or to provide greater than 50\% share of the blockchain's computing power to a malicious group that can undo transactions.
\par This can be avoided by allowing Bitcoin's cryptographic puzzle as an alternate problem that miners can solve to add new blocks. As we want to incentivize solving useful problems, the difficulty of the cryptographic puzzle can be kept high enough so that miners only turn to solve the cryptographic puzzle when there is no useful problem to solve. 
\subsubsection{Deliberately deviating from promised difficulty}
In the centralized scenario, $\mathcal{C}$ can also have their own mining nodes and depending on the fraction of computation power they have, they can increase their rewards earned by controlling difficulty. For example, when the number of miners is less and $\mathcal{C}$ has 25\% of the computation power of the network, they can increase the block publishing rate as compared to when miners are more and $\mathcal{C}$'s power is only 10\% of the network.
\par
It can be argued that any attack by the problem setter is self-sabotage because the network is dedicated to solving their problem. Malicious activity will result in a decrease in the trust in their blockchain, thereby reducing the miners mining in their blockchain and decreasing the network's computation power. Other attacks by hackers can be avoided with good security practices by $\mathcal{C}$.
\subsubsection{Skewed spaces}
In skewed spaces or where alternate and more efficient methods can find preimages for the solution, miners can instead do the following: find the set of ideal inputs and then just hash the block header for different nonces to get $h$ belonging to this ideal set. This will essentially convert to Bitcoin's puzzle but will also give an advantage to the miner who can compute the ideal set quicker. This will only be possible in problems that are poorly designed. Our proposed problem, generating vanity .onion URL, doesn't have this issue.
\subsection{Adoption}
The framework we discussed can be adopted in various ways by different blockchains. We discussed that in BOINC whitelisted projects, contributors are rewarded with Gridcoin \cite{gridc}. BOINC has a leaderboard and scrapers to rank and reward individual contributors. This is disliked by proponents of decentralization as the reported leaderboard can be manipulated to reward their own nodes with more coins. Crowdsourcing projects solving problems that can be adapted to this framework can benefit from a blockchain system where the only role of the central authority will be to set the problem. Verification will be done by the community and all contributors will be fairly rewarded. Such a system will attract both science/maths enthusiasts and miners willing to earn rewards by doing useful work.
\par
This need not be limited to purely computational projects. Programs like Planet Hunters \cite{planethunters} can follow the same model. Citizen Scientists can fetch the data determined by the hash of their block and do the manual work.
\par
We also showed how problem setting can be decentralized by allowing anyone to propose a problem and offer a problem fee. This idea can be adopted by existing proof-of-work blockchains or be started as a new blockchain \vc. The subset $\mathcal{S}$ can vary depending on the problems the blockchain developers promise to focus on. If a newer problem $\mathcal{P}_i \in \mathcal{S}_s$ is discovered later, it can be added to the set $\mathcal{S}$ for the blockchain if the network agrees to.

\section{Acknowledgement} \label{acknowledgement}
The research was partially funded through National Unified Blockchain Framework by MEITY, India.


\section{Conclusion} \label{conclusion}
We proposed a formal framework, \ourname, to show how proof-of-work protocols can be modified to do
practically 
useful work. 
We discussed both centralized and decentralized scenarios to allow both crowdsourcing organisations and individual puzzlers to get their problems solved.
We also proposed one useful problem type that can replace the partial hash pre-image cryptographic puzzle, thereby better utilizing the energy spent on mining. This resulted in \vc, a blockchain devoted to generating vanity addresses for \emph{\puzzlers} as proof-of-work.
We believe, to make blockchains more sustainable, \ourname\ is going to be very helpful.

\bibliographystyle{plain}
\bibliography{references}

\section{Appendix} \label{appendix}
\subsection{Using Ed25519 for address generation}
The following is the brief explanation of how Ed25519 is used for key pair generation \cite{ed_add} :
\begin{itemize}
    \item A 256-bit random seed is taken as input to SHA512, giving a 512 bit digest.
    \item Only the left half $l$, i.e. the left 256 bits of the 512 bit digest is used for public key generation and is treated as 32-byte number.
    \item Pruning: The lowest three bits of the first byte and highest bit of the last byte are made 0 and the second highest bit of the last byte is set to 1.
    \item This 32-byte buffer is now interpreted as little-endian integer. Let's denote this secret scalar as $s$.
    \item $B$ is the base point of the curve. $\mathcal{A} = [s]B$. This $\mathcal{A}$ is the public key (after some bit manipulation) \cite{ed_wiki}.
\end{itemize}
From this public key $\mathcal{A}$, the onion address is generated as follows\cite{rend-spec}:
\begin{itemize}
    \item Checksum $C_s = \mathcal{H}$(".onion checksum" $|| \mathcal{A} ||$ version)[:2], i.e. hash of the string ".onion address" appended with public key and version number truncated to 2 bytes.
    \item version is one byte version field.
    \item Onion address $\mathcal{O} =$ base32($\mathcal{A} ||$ $C_s ||$  version ) + ".onion"
\end{itemize}
\subsection{Adapting Ed25519 for use as PoUW problem}
Problem Proposal:
\begin{itemize}
    \item Let the \puzzler \ select their own left half $l_0$ and right half of the 512-bit secret key such that the left half satisfies the pruning step, i.e. lowest three bits of the first byte and highest bit of the last byte should be 0 and second highest bit of the last byte should be 1.
    \item We additionally require third bit of the last byte to be 0. The left half is interpreted as little-endian integer $s_0$.
    \item The \puzzler \ publishes $\mathcal{A}_0 = [s_0]B$ with the valid regex $\mathcal{R}$. 
\end{itemize}
Mining:
\begin{itemize}
    \item The hash $h$ of the block header is of 256-bit.  The last 27 bits are set to 0. The first 27 bits are set to 0 \footnote{We do this step so that the resulting $s$ comes from a valid $l$ after pruning. Setting 27 bits to 0 is not necessary as only the appropriate bits can be set to zero but we do so for simplicity.} \footnote{Note that we have already taken into account the little-endian interpretation}. Denote this as $h'$. 
    \item Calculate $\mathcal{A} = \mathcal{A}_0 + [h']B$.
    \item If $\mathcal{A} \in \mathcal{R}$, i.e., if the resulting public key satisfies the regex description, the problem is considered solved.
    \item The miner repeats this process with a different nonce if $\mathcal{A} \not \in \mathcal{R}$.
\end{itemize}
Using the resulting public key as the address:
\begin{itemize}
    \item The \puzzler \ can get back left half $l'$ corresponding to $h'$ by interpreting $h'$ as little-endian.
    \item $\mathcal{A} = [s_0]B + [h']B = [s_0 + h']B$. Left half $l = l_0 + l'$. 
    \item Onion address $\mathcal{O}$ is then generated by the same procedure.
\end{itemize}
As $l_0$ and $s_0$ are only known to the \puzzler, $l$ and $s$ can only be computed by the \puzzler, thus getting the secure vanity .onion address $\mathcal{O}$ they wanted.

\subsection{Difficulty of VanityCoin Problem Proposal}
Total number of possible TOR v3 addresses $\sim 32^{56}$ and TOR v2 addresses $\sim 32^{16}$. RegEx for the entire (v2) space:\\ 
$$\wedge([a-z2-7])\{16\}\$$$\\
\hspace*{1em} If a puzzler wants domain names either \emph{i) crypto**********.onion or ii) coin************.onion}, the regex $\mathcal{R}$ would be: \\
$$\wedge(c(rypto([a-z2-7]\{10\})|oin([a-z2-7]\{12\})))\$$$\\
\hspace*{1em}Probability of random URL satisfying this: $1/32^4 + 1/32^6 \approx 9.546 * 10^{-7}$. Using the hash rate of the blockchain and the average block publishing rate, the required probability for $\mathcal{R}$ can be varied.

\subsection{Abbreviations and Notations}
\begin{tabular}{m{2cm}|m{13cm}}
    \hline
    PoW & Proof of Work \\
    PoUW & Proof of Useful Work\\
    PUPoW & Practically Useful Proof of Work \\
    BOINC & Berkeley Open Infrastructure for Network Computing \\
    ASIC & Application-Specific Integrated Circuit \\
    DSA & Digital Signature Algorithm\\
    Ed25519 & Edwards-curve Digital Signature Scheme using SHA-512 and Curve25519. \\
    Tor & The Onion Router \\
    $\mathcal{H}$ & Hashing function of the required length \\
    $\mathcal{C}$ & Centralized Problem Setter \\
    $\mathcal{S}_s$ & Set of all problem types satisfying PUPoW problem requirements \\
    $\mathcal{S}$ & Subset of $\mathcal{S}_s$ which the blockchain chooses to allow for valid problem proposals. \\
    $\mathcal{R}$ & Regular expression specified in the problem proposal. \\
    \hline
\end{tabular}
\normalsize

\end{document}